\documentstyle[amsfonts]{article}
\newcommand{\bu}{1\,\!\rule{0.2mm}{3.1mm}
\hspace{-1.3mm}\rule[3.1mm]{1.6mm}{0.1mm}
\hspace{-1mm}\rule{1.1mm}{0.1mm}}
\title{Quantum inconsistency of $W_3$-gravity models\\
associated with magical Jordan algebras}
\author{S. A. Lyakhovich\thanks{e-mail:SSL@fftgu.tomsk.su}
and A. A. Sharapov}
\date{Physics Department, Tomsk State University,\\
Tomsk 634050, Russia}
\begin{document}
\Large

\maketitle
\begin{abstract}
It is shown that the Sugawara-type construction for $W_3$ algebra
associated with the four magical Jordan algebras leads to the anomalous
theory of $W_3$ gravity.
\end{abstract}

\newpage

The Virasoro algebra plays a key role in the study of string models and
$2D$-conformal field theory, with various extensions involving
Kac--Moody or superconformal generators. The $W_3$ algebra, proposed
by Zamolodchikov [1] as a non-linear extension of the Virasoro algebra,
has provided a basis for generalization of all $2D$-conformal field theory
models to the case more wide and non-linear symmetry. In particular
these are so-called the models of $W_3$ gravity [2, 3] and the $W_3$
strings [4, 5]. The $W_3$ Zamolodchikov's algebra underlying these
models is described by the following operator products:
$$
\hbar^{-1}T(z)T(w) \sim \hbar \frac{d/2}{(z-w)^4} + \frac{2T(w)}
{(z-w)^2} + \frac{\partial T(w)}{z-w};
$$
$$
\hbar^{-1}T(z)W(w) \sim \frac{3W(w)}{(z-w)^2} + \frac{\partial W(w)}{z-w};
\eqno{(1)}$$
$$
\hbar^{-1}W(z)W(w) \sim \hbar^2 \frac{d/3}{(z-w)^6} + \hbar\frac{2T(w)}
{(z-w)^4} + \hbar\frac{\partial T(w)}{(z-w)^3} +
$$
$$+ \hbar\frac{3}{10}~ \frac{\partial^2 T(w)}{(z-w)^2} + \hbar\frac{1}{15}~
\frac{\partial^3 T(w)}{z-w} + \frac{2\beta\Lambda (w)}{(z-w)^2} +
\frac{\beta\partial \Lambda (w)}{z-w}.
$$
The constant $\beta$ is determined by the associativity requirement to
take the value
$$
\beta = \frac{16}{22 + 5d},
\eqno{(2)}$$
and $\Lambda (z)$, which includes the non-linearities referred to
above, is a composite field formed from the Virasoro generators [6]:
$$
\Lambda(w) = \oint\frac{{\rm d}z}{z-w}T(z)T(w) - \frac{3}{10}\hbar
\partial^2T(w).
\eqno{(3)}$$

The general free-field ansatz for the $W_3$ generators in terms of the
fundamental spin-one currents $J^i$ reads [2]
$$
\begin{array}{l}
T = -~\frac{\displaystyle 1}{\displaystyle 4}g_{ij}:J^iJ^j:+{\rm
i}\hbar a_j\partial J^j,\vspace{2\baselineskip} \\
W = -~\frac{\displaystyle{\rm i}}{\displaystyle 12}d_{ijk} : J^iJ^jJ^k
: - \sqrt{\hbar}e_{ij}:J^i\partial J^j:+{\rm i}\hbar f_j\partial^2J^j.
\end{array}
\eqno{(4)}$$
The notation : : indicates normal ordering with respect to the modes
of\linebreak
$J^i = \sum J^i_nz^{n-1}$ which have conventional operator products
$$
J^i(z)J^j(w) \sim \frac{\hbar g^{ij}}{(z-w)^2}.
\eqno{(5)}$$
The leading terms in Eq. (4) directly correspond to the classical
currents appearing in $W_3$ gravity.

In order to hold the $W_3$ algebra for the above realization $T$ and
$W$ some set of conditions must be imposed on the structure tensors
$g_{ij}$, $d_{ijk}$, $e_{ij}$, $f_i$, $a_i$. ${}$ In classical limit ($\hbar
\to 0$) these conditions imply that $d_{ijk}$ must
necessarily be a structure constant of some Jordan algebra ${\Bbb J}_3$
with a cubic norm [7, 8]. The full set of constraints on the structure
tensors as well as their solutions for the cases of generic Jordan
algebras was formulated in ref. [9]. However the solution existence
problem still remains unsolved for the other four cases known as the
magical Jordan algebras ${\Bbb J}^{\Bbb R}_3$, ${\Bbb J}^{\Bbb C}_3$,
${\Bbb J}^{\Bbb H}_3$, ${\Bbb J}^{\Bbb O}_3$ with dimensions 5, 8, 14
and 26 respectively. The set of constraints on the structure tensors
corresponding to the magical Jordan algebras takes the form
$$
\begin{array}{c}
a_i = f_i = 0, \qquad g^{ij}d_{ijk} = 0; \\
d^m_{(ij}d_{k)ml} = \mu^2g_{(ij}g_{kl)}; \\
e_{ij} = -e_{ji}, \qquad {e_i}^je_{jk} = -\rho^2g_{ik}; \\
{e_i}^ld_{jkl} = {e_j}^ld_{ikl}, \end{array}
\eqno{(6)}$$
where all indices are raised with $g^{ij}$-inverse matrix to $g_{ij}$
and $i=1,{\dots},n$, $n=5,8,14,26,$
$$
\mu^2=\frac{16}{22+5n}, \qquad \rho^2=\frac{n-2}{8(22+5n)}
$$
In present paper we will prove the statement that the set of equations
(6) has no solutions for any magical Jordan algebra. This fact makes
impossible to construct consistent quantum theory of $W_3$ gravity
models considered, as it is shown in the paper below.

First of all we note that making redefinition $d_{ijk} \to
\frac{1}{\mu}d_{ijk}$, $e_{ij} \to \frac{1}{\rho}e_{ij}$ in equations
(6) one can put $\rho = \mu = 1$. For purposes of following
calculations it is convenient to introduce matrices
$$
D({\bf p}) = \|p^kd^{~i}_{k~j}\|, \qquad M({\bf q}, {\bf p}) =
\|q^ip_j + p^iq_j\|, \qquad E = \|{e_i}^j\|
\eqno{(7)}$$
where ${\bf p}$ and ${\bf q}$ are arbitrary vectors. Then equations
(6) can be rewritten in the matrix form
$$
{\rm Sp}~D({\bf p}) = 0, \qquad E^{\rm T} = -E, \qquad E^2 = -{\bu};
$$
$$
\{D({\bf p}), D({\bf q})\} = -D(D({\bf p}){\bf q})
+ M({\bf p},{\bf q}) + ({\bf p}, {\bf q}){\bu};
\eqno{(8)}$$
$$
ED({\bf p}) = D({\bf p})E^{\rm T}.
$$
Here brackets \{ , \} stand for a matrix anticommutator, and $({\bf
p}, {\bf q}) = p_iq^i$ is inner product of two vectors.

It is easy to see, that matrices $D({\bf r})$, $M({\bf p}, {\bf q})$
and ${\bu}$ generate a closed algebra with respect to their
anticommutator, i.e. represent some Jordan algebra $\tilde{\Bbb
J}$\footnote{It is interesting to note that the matrices $\{M\}$
generate an ideal of the $\tilde{\Bbb J}$, and associated factor-algebra
$\tilde{\Bbb J}/\{M\}$ appears to be isomorphic to the initial
${\Bbb J}_3^{({\Bbb R},{\Bbb C},{\Bbb H},{\Bbb O})}$.}.
Indeed
$$
\begin{array}{c}
\{ {\bu}, M({\bf p}, {\bf q})\} = 2M({\bf p}, {\bf q}), \qquad \{
{\bu}, D({\bf p})\} = 2D({\bf p}),\vspace{2\baselineskip} \\
\{D({\bf r}), M({\bf p},{\bf q})\} = M(D({\bf r}){\bf p},{\bf q})
+ M({\bf p}, D({\bf r}) {\bf q}),\vspace{2\baselineskip} \\
\{M({\bf a},{\bf b}),M({\bf p},{\bf q})\}=({\bf a},{\bf p})M({\bf b},
{\bf q})+({\bf a},{\bf q})M({\bf b},{\bf p})+\vspace{2\baselineskip}\\
+({\bf b},{\bf q})M({\bf a},{\bf p})+({\bf b},{\bf p})M({\bf a},{\bf q}).
\end{array}
\eqno{(9)}$$
The above matrices have the following values of traces:
$$
{\rm Sp}~D({\bf p}) = 0, \qquad {\rm Sp}~M({\bf p},{\bf q}) =
2({\bf p},{\bf q}), \qquad {\rm Sp}~{\bu} = n.
\eqno{(10)}$$
The next matrix relation immediately follows from Eq. (8)
$$
D({\bf p}) = -ED({\bf p})E^{\rm T}.
\eqno{(11)}$$
Razing to an odd power both side of Eq. (11) and computing traces of
derived expressions we see
$$
{\rm Sp}~(D({\bf p}))^{2m+1} = 0, \quad m = 1, 2, \ldots ~.
\eqno{(12)}$$
On the other hand, using relation
$$
{\rm Sp}~D^{n+1} = \frac{1}{2^n}{\rm Sp}~\underbrace{\{\cdots\{\{}_{n}
D,D\},D\},\cdots\}
\eqno{(13)}$$
algebra (8), (9) and expressions for traces (10) one can sequentially
obtain trace of any power of $D({\bf p})$. For a few lower powers of
the matrix $D({\bf p})$ we get
$$
{\rm Sp}~D^2({\bf p}) = \frac{2+n}{2}{\bf p}^2, \qquad {\rm
Sp}~D^4({\bf p}) = \frac{10+3n} {8}({\bf p}^2)^2,
$$
$$
{\rm Sp}~D^3({\bf p}) = \frac{2-n}{4}{\bf p}^3, \qquad {\rm
Sp}~D^5({\bf p}) = \frac{2-5n}{16}{\bf p}^2{\bf p}^3,
\eqno{(14)}$$
$$
p^2 \equiv p_ip^i, \qquad p^3 \equiv d_{ijk}p^ip^jp^k.
$$
Comparing Eqs. (14) with Eq. (12) we come to contradiction.

{}From the above discussion it appears that the classical $W_3$ algebra
associated with magical Jordan algebras can not be extended to the
quantum level by modifying classical currents by terms proportional to
$\sqrt{\hbar}$ and $\hbar$. The quantum selfconsistency is known to
require the BRST charge operator to be nilpotent. The quantum BRST
charge for Zamolodchikov algebra has the form [10, 11]
$$
\Omega = \oint{\rm d}z :\{c(T - {\bar c}c' - \frac{1}{2}{\bar c}'c -
3{\bar b}b' - 2{\bar b}'b) - bW - \beta b{\bar c}b'T - \frac{5d}{1044}
(\frac{1}{15}{\bar c}''b'b - \frac{1}{6}{\bar c}b''b')\} :.
\eqno{(15)}$$
Here $c(z)$ and ${\bar c}(z)$ are a pair of conformal ghosts, while
$b(z)$ and ${\bar b}(z)$ are a new pair of ghosts related with $W$.
The requirement of nilpotency $\Omega^2=0$ fixes the critical value of
$d= 100$. Since the algebra of $W_3$ symmetry is unclosed in the quantum
level the $\Omega$ charge is not nilpotent and this as well known gives
rise to anomalies. Our result supplements the findings of ref. [9] and
makes possible to assert that all anomaly-free models of $W_3$
gravity one--to--one correspond to the generic Jordan algebras ${\Bbb
J}_3$.

This work supported in part by European Community Commission contract
INTAS-93-2058 and by the International Science Foundation grant No
M2I000.

\newpage

\centerline{References}

\bigskip

\noindent
1. A.B. Zamolodchikov, Teor. Mat. Fiz., {\bf 65}, 374 (1985).\\
2. C.M. Hull, Phys. Lett. B, {\bf 240}, 110 (1990).\\
3. K. Schoutens, A. Sevrin and P. van Nieuwenhuizen, Phys. Lett. B,
{\bf 243}, 245 (1990).\\
4. C.N. Pope, L.J. Romans and K.S. Stelle, Phys. Lett. B, {\bf 269},
287 (1991).\\
5. C.N. Pope, L.J. Romans, E. Sezgin and K.S. Stelle, Phys. Lett. B,
{\bf 274}, 298 (1992).\\
6. B.L. Feigin and D.B. Fuchs, Funktc. Anal. Pril., {\bf 16}, 47 (1982).\\
7. C.M. Hull, Nucl. Phys. B, {\bf 353}, 707 (1991).\\
8. M. G\"unuydin, G. Sierra and P.K. Townsend, Nucl. Phys. B,
{\bf 242}, 244 (1984).\\
9. L.J. Romans, Nucl. Phys. B, {\bf 352}, 829 (1991).\\
10. J. Thierry-Mieg, Phys. Lett. B, {\bf 197}, 368 (1987).\\
11. Won-Sang Chung and Jae-Kwan Kim, Phys. Rev. D, {\bf 41}, 1336
(1990).
\end{document}